\documentclass[12pt]{article}
\usepackage{indentfirst}
\usepackage{graphicx}

\title{Laser cooling of unbound atoms in nondissipative optical
lattices}
\author{ N.A.Matveeva, A.V.Taichenachev, A.M.Tumaikin, V.I.Yudin\\
{\small Novosibirsk State
University, Novosibirsk 630090,Russia}\\
{\small Institute of Laser Physics SB RAS, Novosibirsk 630090,
Russia}}\begin{document}

\maketitle The semiclassical theory of laser cooling is applied for
the analysis of cooling of unbound atoms with the values of the
ground and exited state angular moments $1/2$ in a one-dimensional
nondissipative optical lattice. We show that in the low-saturation
limit with respect to the pumping field a qualitative interpretation
of the cooling mechanisms can be made by the consideration of
effective two-level system of the ground-state sublevels. It is
clarified that in the limit of weak Raman transitions the cooling
mechanism is similar to the Doppler mechanism, which is known in the
theory of two-level atom. In the limit of strong Raman transitions
the cooling mechanism is similar to the known Sisyphus mechanism. In
the slow atom approximation the analytical expressions for the
coefficients of friction, spontaneous and induced diffusion are
given, and the kinetic temperature is estimated.

 PACS number(s):
32.80.Pj,32.80.Lg,39.25.+k,39.10.+j
\section{Introduction}


Laser cooling of neutral atoms is necessary in various fundamental
and applied problems, such as high precision spectroscopy
\cite{spectroscopy}, atomic frequency standards \cite{standarts1,
standarts2, standarts3}, Bose-Einstein condensation \cite{condens},
atomic nanolitography \cite{litogr1, litogr2} and others. The
methods of cooling of neutral atoms in magneto-optical traps and
optical molasses that give the temperature of atomic ensemble about
$\mu$K have been developed for he last 20 years. However, lower
temperatures are required for some applications. Particularly, the
sub-$\mu$K transversal cooling would allow one to achieve the higher
precision and stability in the modern laser-cooled atomic frequency
standards(the atomic fountains \cite{standarts2, standarts3}, the
atomic clock in the condition of microgravitation \cite{clock} ). At
the present time there exist several methods of laser cooling, which
allow one to achieve the temperature of atomic ensemble below
$\mu$K: the velocity-selective coherent population trapping
\cite{KPN},the cooling by Raman pulses \cite{ramanpuls,2Draman} and
the degenerate sideband Raman laser cooling (further DSRLC )
\cite{jessen, chy}. DSRLC appears to be an adaptation of the
resolved-sideband laser cooling of ions for neutral atoms
\cite{ions}. In comparison with the other methods of laser cooling
DSRLC has some advantages: high efficiency, relatively short cooling
time (about ms), relative simplicity of experimental realization.
This method is based on the use of the Raman two-photon transitions
between the vibrational levels of the Zeeman substates of atoms,
that are trapped in an optical lattice. In the paper \cite{jessen}
the experiments on the two-dimensional cooling of cesium atoms by
this method up to the ground state of a far-resonance optical
lattice have been reported. In these experiments the DRSLC stage was
preceded by the precooling stage in a near-off-resonance optical
lattice, that ensured the high efficiency of cooling ($95\%$ of
atoms that were captured in magneto-optical trap were cooled up to
the ground vibrational state of lattice), but it caused some
complications of the experimental realization. Chu with co-authors
carried out similar experiments on three-dimensional cooling of
cesium atoms in optical lattice up to the kinetic temperature 290 nK
(after the adiabatic release of atoms from a lattice \cite{chy}).
The distinction of this experiments from \cite{jessen} appears to be
the absence of the  precooling stage. Nevertheless, 80 \(\%\) of
atoms, that are transferred in a three-dimensional lattice, are
cooled up to their ground vibrational state.

 The high cooling efficiency
which has been achieved \cite{chy}, was most likely the evidence of
the co-existence of cooling mechanisms of bound and unbound atoms
that was shortly discussed in \cite{chy}. Later the experiments on
2D laser collimation of a continuous beam of cold cesium atoms by
the method of DSRLC \cite{swid} were carried out to improve the
corresponding frequency standard. In these investigations the
cooling scheme similar to that in \cite{swid} is employed, but with
some distinctions, that, particularly, lie in the use of a
two-dimensional optical lattice of the original configuration.
However, the efficiency of transversal cooling (collimation) of
atomic beam is not high enough and it was essentially lower in
comparison with \cite{chy}. The reasons of the lower cooling
efficiency have not been investigated in \cite{swid}.

Therefore the necessity of more detailed investigation of cooling in
a nondissipative optical lattices arises. Particularly, it is
important for the revealing of conditions, when the co-existence of
cooling mechanisms of bound and unbound atoms takes place. In the
present paper the semiclassical theory of laser cooling is used for
the analysis of cooling of unbound atoms in a nondissipative optical
lattice. This analysis is made in the framework of the simplest
model of atoms with the degenerate ground state and for
one-dimensional configuration of the lattice field that reflect
essential features of the experimental scheme \cite{swid}. We
consider the one-dimensional atomic motion, neglecting the recoil in
all other directions.

As a result, the qualitative interpretation of the cooling
mechanisms is given, and  the analytical expressions for the force
acting on an atom , the coefficients of spontaneous and induced
diffusion are obtained. This allows the quantitative estimations of
the atomic kinetic parameters, particularly, of the temperature.
\section{Statement of the problem}

\begin{figure}[t]
\begin{center}
\includegraphics[height=6cm]{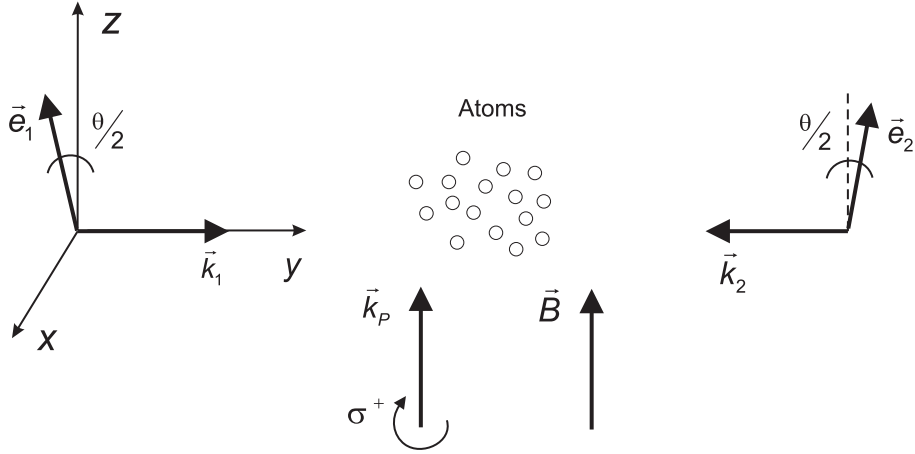}
\caption{Geometry of task}\label{fig:config}
\end{center}
\end{figure}

Let us consider a two-level atom with the angular momentum of the
ground state \(J_{g}=1/2\) and the momentum of the exited state
\(J_{e}=1/2\),moving in an optical lattice. The lattice field is
formed by two counterpropagating (running along $y$ axis) linearly
polarized laser beams, their polarization vectors \(\textbf{e}_{1}\)
and \(\textbf{e}_{2}\) are directed with angle \(\theta\) to each
other making \textit{lin-\(\theta\)-lin} configuration
(fig.\ref{fig:config}).

\begin{figure}[t]
\centering
\includegraphics[height=6cm]{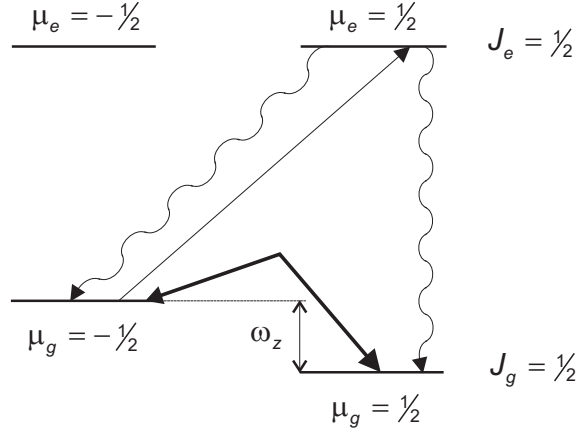}
\caption{Scheme of transitions. The thick double arrow labels Raman
two-photon transition under the lattice field action, the thin arrow
labels transition under the pumping field action, the wavy arrows
label the spontaneous decay of exited state. }\label{fig:transit}
\end{figure}

In the spherical basis the lattice field has the following form:
\begin{equation}
\textbf{E}_{L}(y)=E_{0}^{L}\exp(-i\omega_{L}\label{eq:lattice}
t)\sum_{q=0,\pm1}e_{L}^{q}(y)\textbf{e}_{q},
\end{equation}
where \(E_{0}^{L}\) is the amplitude of single beam,
\(\textbf{e}_{q}\) are spherical orts. Let polarization vectors
\(\textbf{e}_{1}\) and \(\textbf{e}_{2}\) are directed with the
angle \(\theta/2\) to the axis of the quantization (axis $z$ on
fig.\ref{fig:config}), then the contravariant components
\(e^{q}_{L}(y)\) are written as:
\begin{eqnarray}
e_{L}^{0}&=&2\cos(\theta/2)\cos(ky),\nonumber\\
e_{L}^{-1}&=&\sqrt{2}i\sin(\theta/2)\sin(ky),\nonumber\\\label{eq:complat}
e_{L}^{+1}&=&-\sqrt{2}i\sin(\theta/2)\sin(ky).
\end{eqnarray}
It is assumed that the lattice field is detuned far enough from the
resonance:\(|\delta_{L}|>>\gamma\) ,
(\(\delta_{L}=\omega_{L}-\omega_{0}\) is the detuning of the lattice
field frequency $\omega_{L}$ from the frequency of atomic transition
$\omega_{0}$, \(\gamma\) is the relaxation rate of the exited
state), so that we can neglect the real transitions of atom from the
ground state to the exited state under the action of lattice field.
Since the spontaneous emission of photons is also negligibly small,
such a lattice is nondissipative. So, the lattice action comes to
the forming of periodical potential and the inducing of Raman
two-photon transitions between Zeeman sublevels of the ground state
(on fig.\ref{fig:transit} these transitions are labeled with the
thick double arrow).

The lattice field action on atoms alone is not enough for cooling,
because the lattice is nondissipative, that is the atomic motion in
such a lattice has the conservative character.
 It is necessary for
the realization of cooling in this system the presence of pumping
field, which is tuned to resonance with the atomic transition, and
spatially uniform splitting of the ground state Zeeman sublevels.
The pumping field represents the circularly polarized beam, which is
directed along the axis $z$:
\begin{equation}
\textbf{E}_{p}=E_{0}^{p}\exp(-i \omega_{p}\label{eq:pump}
t)\exp(ikz)\textbf{e}_{+1},
\end{equation}
where \(E_{0}^{p}\)  is the pumping field amplitude The resonant
pumping field induces the one-photon transitions from the ground
state sublevel with the projection \(-1/2\) to the exited state
sublevel with the projection \(1/2\) (the thin arrow on fig.
\ref{fig:transit} ). Further the spontaneous decay of exited state
occurs (the wavy arrows on fig.\ref{fig:transit} ). We neglect the
recoil effect pumping field, because we consider cooling only in the
$y$ direction. So, the pumping field action (together with the
spontaneous decay from the exited state) leads to the effective
relaxation of the ground state sublevels system. The spatially
uniform shift of the Zeeman sublevels is produced by a static
magnetic field (Zeeman effect), its direction coincides with the
direction of pumping field wave vector. We do not take into account
the splitting of the exited state Zeeman sublevels, considering it
to be much smaller of the exited state natural width. Cooling in
this system can be achieved by a proper choice of the ground state
Zeeman splitting magnitude.

The evolution of atomic system is described by the quantum kinetic
equation (QKE) for the atomic density matrix. In our case, in the
general form, without the concretization of representation, QKE can
be written as:
\begin{eqnarray}
\frac{d\widehat{\rho}}{dt}=-\frac{i}{\hbar}\left[\widehat{H}_{0}+\frac{\widehat{p}_{y}^{2}}{2M},\widehat{\rho}\right]-\nonumber\\
-\frac{i}{\hbar}\left[\widehat{V}_{L}+\widehat{V}_{P}+\widehat{V}_{B},\widehat{\rho}\right]-\widehat{\Gamma}\{\widehat{\rho}\}.\label{eq:KKE}
\end{eqnarray}
Here \(\widehat{p}_{y}\) is the operator of atomic momentum
projection on the axis $y$, \(\widehat{H}_{0}\) is the Hamiltonian
of free atom in the rest:
\begin{equation}
\widehat{H}_{0}=\hbar\omega_{0}\sum_{\mu_{e}}\left|J_{e}\mu_{e}\right>\left<J_{e}\mu_{e}\right|,\label{eq:H0}
\end{equation}
where \(\omega_{0}\) is the atomic transition frequency, \(J_{e}\)
is the angular momentum of the exited state, \(\mu_{e}\) is its
projection on the quantization axis. The operator of interaction of
atom with lattice field has the form:
\begin{equation}
\widehat{V}_{L}(y)=\hbar\Omega_{L}\sum_{q}\widehat{T}_{q}e_{L}^q(y)\exp(-i\omega_{L}t)+h.c.,\label{eq:VL}
\end{equation}
where \(\Omega_{L}=-\frac{\widetilde{d}E_{0}^{L}}{\hbar}\) is the
Rabi frequency per a single beam of lattice field,
(\(\widetilde{d}\) is the reduced matrix element of the dipole
moment operator). According to the Wigner-Eckart \cite{varsh}
theorem the dependence of operator $\widehat{V}_{L}(y)$ on the
magnetic quantum numbers is contained in the Wigner operator:
\begin{equation}
\widehat{T}_{q}=\sum_{\mu_{e}\mu_{g}}\left|J_{e}\mu_{e}\right>C_{J_{g}\mu_{g}1q}^{J_{e}\mu_{e}}\left<J_{g}\mu_{g}\right|,\label{eq:vign}
\end{equation}
where \(J_{g}\) and \(\mu_{g}\) is the ground-state angular momentum
and its projection, \(C^{J_{e}\mu_{e}}_{J_{g}\mu_{g}1q}\) is the
Clebsch-Gordan coefficients. The operator of interaction of atom
with pumping field is analogously written as:
\begin{equation}
\widehat{V}_{p}=\hbar\Omega_{p}\widehat{T}_{+1}\exp{(ikz)}\exp(-i\omega_{p}t)+h.c.,\label{eq:Vp}
\end{equation}
where \(\Omega_{p}\) is the pumping field Rabi frequency. The
operator of interaction of atom with magnetic field can be written,
taking into account only the linear Zeeman effect in the ground
state:
\begin{equation}
\widehat{V}_{B}=-\hbar\omega_{z}\widehat{J}_{gz},\label{eq:Vb}
\end{equation}
where $\omega_{z}$ is the Zeeman splitting of ground state
sublevels, $\widehat{J}_{gz}$ is the operator of $z$-projection
ground-state angular momentum. The action of the atomic radiation
relaxation operator \(\widehat{\Gamma}\{\widehat{\rho}\}\) can be
presented as:
\begin{eqnarray}
\widehat{\Gamma}\{\widehat{\rho} \}
=\frac{\gamma}{2}\{\widehat{P}_{e},\widehat{\rho}\}-\gamma\frac{3}{2}\left\langle\sum_{s=1,2}(\widehat{\textbf{T}}\cdot\textbf{e}_{s}(\textbf{k}))^{\dag}\right.\\
\left.\exp(-i k_{y}\widehat{y})\widehat{\rho} \exp(i
k_{y}\widehat{y})(\widehat{\textbf{T}}\cdot\textbf{e}_{s}(\textbf{k}))\right\rangle_{\Omega_{k}},\label{eq:Gamma}
\end{eqnarray}
where \(\widehat{P_{e}}\) is the projector on the exited state:
\begin{equation}
\widehat{P}_{e}=\sum_{\mu_{e}}\left|J_{e}\mu_{e}\right>\left<J_{e}\mu_{e}\right|,\label{eq:Pe}
\end{equation}
\(\textbf{k}\) is the wave vector of spontaneous photon,
\(\textbf{e}_{s}(\textbf{k})\) is the unit polarization vectors of
spontaneous photon, which are orthogonal \(\textbf{k}\), \(<\ldots
>_{\Omega_{k}} \) denotes averaging on running direction of spontaneous photons, \(k_{y}=(\textbf{k}\cdot\textbf{e}_{y})\).

As is well known \cite{forse}, one of the conditions for
quasiclassical atomic translation motion is a smallness of recoil
parameter, which is the ratio of photon momentum \(\hbar k\) to
atomic momentum dispersion \(\triangle p\):
\begin{equation}
\frac{\hbar k}{\triangle p}<<1.\label{eq:otdacha}
\end{equation}
The execution of condition (\ref{eq:otdacha}) allows one to separate
the fast process of the ordering on internal degrees of freedom from
slow processes, which are connected with translation motion. At the
kinetic evolution stage (in our case at \(t>>(\gamma S_{p})^{-1})\),
when the stationary distribution on internal degrees of freedom has
established, the atomic ensemble dynamics is defined by slow
processes of change of distribution function on translation degrees
of freedom. Usually the Wigner representation is used for the
translation degrees of freedom, then the initial QKE (\ref{eq:KKE})
is reduced (with account for the terms of second order in the recoil
parameter ) to the closed equation of the Fokker-Plank type for the
Wigner distribution function \(W\) :

\begin{equation}
\left(\frac{\partial}{\partial
t}+\frac{p_{y}}{M}\frac{\partial}{\partial
y}\right)W(y,p_{y})=\left[-\frac{\partial F(y,p_{y})}{\partial
p_{y}}+\frac{\partial^{2}D(y,p_{y})}{\partial
p_{y}^{2}}\right]W(y,p_{y}).\label{eq:Foker}
\end{equation}
The function \(W\) can be interpreted as the probability density in
the phase space in the case \(W\) is a positive definite. The
coefficients \(F(y,p_{y})\) and \(D(y,p_{y})\) have meaning of the
force and diffusion in the momentum space, respectively. The
coefficient of diffusion \(D(y,p_{y})\) is presented in the form:
\begin{equation}
D(y,p_{y})=D_{sp}(y,p_{y})+D_{ind}(y,p_{y}),\label{eq:D}
\end{equation}
where \(D_{sp}(y,p_{y})\) is the coefficient of spontaneous
diffusion, \(D_{ind}(y,p_{y})\) is the coefficient of induced
diffusion \cite{minogin}.

\section{Effective two-level system}
For a qualitative interpretation of cooling in the system under
consideration it is appropriately to use the approximation of low
saturation by pumping field, that corresponds to the experimental
conditions \cite{swid}. This condition can be written as:
\begin{equation}
S_{p}=\frac{\Omega_{p}^{2}}{(\gamma/2)^{2}+\delta_{p}^{2}}<<1,\label{eq:sut}
\end{equation}
where \(S_{p}\) is the saturation parameter for pumping field,
\(\delta_{p}\) is the pumping field detuning from resonance. When
the condition (\ref{eq:sut}) is satisfied,the atomic model under
consideration is equivalent to the two-level ground state substates
system. Really, in this case in the equation (\ref{eq:KKE}) the
standard reduction procedure to the ground state \cite{forse} can be
made. The obtained equation system for the ground-state density
matrix can be compared with the well-known equations for a two-level
atom \cite{minogin}. It is obvious from this comparison, that the
ground state sublevels system is equivalent to the effective
two-level system, where the ground state sublevels with the momentum
projection \(\pm1/2\) play a part of the ground and exited state,
consequently. At that the effective two-level system parameters are
expressed through the initial model parameters as:
\begin{eqnarray}
\Gamma_{1}&=&2/9\gamma S_{p},\nonumber\\
 \Gamma_{2}&=&3\Gamma_{1},\nonumber\\
\Delta&=&-2/3\delta_{p}S_{p}-\omega_{z},\nonumber\\
\chi&=&\frac{2\Omega_{L}^2\sin\theta
}{3\delta_{L}}.\label{eq:effpar}
\end{eqnarray}
Here \(\Gamma_{1}\) is the effective relaxation rate of populations,
\(\Gamma_{2}\) is the effective relaxation rate of coherence,
\(\Delta\) is the effective detuning from the two-photon resonance,
\(\chi\) is the effective Rabi frequency. As is mentioned above,
that in the limit (\ref{eq:sut}) the atomic model under
consideration corresponds to the effective two-level  system. This
fact lies in the base of the qualitative interpretation of cooling
mechanisms.

\section{Qualitative interpretation of cooling mechanisms}
Our analysis of cooling mechanisms in the ground state sublevels
system is based on the well-known cooling mechanisms in the
two-level system. They are the Doppler mechanism in the weak field
limit and the Sisyphus mechanism in the strong field limit
\cite{cazan}. At that together with the conservation of basic
properties of these mechanisms some specific features appear. They
are connected with the two-photon character of excitation and
two-step character of relaxation in the effective two-level system.

\begin{figure}[t]
\centering
\includegraphics[height=6cm]{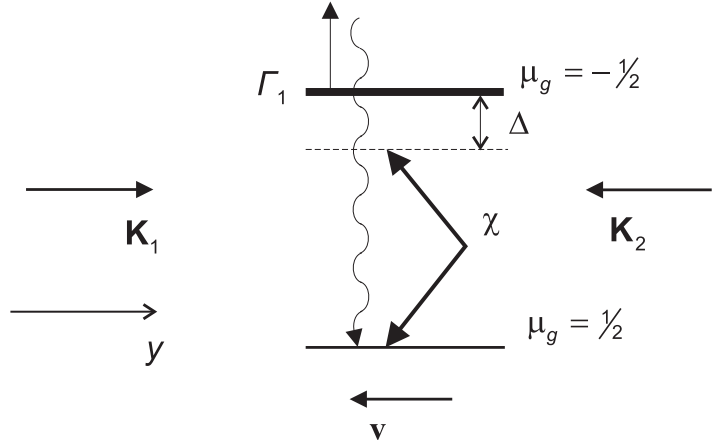}
\caption{Doppler mechanism in two-level effective system
}\label{fig:dopl}
\end{figure}

In the weak Raman transitions limit
$|\chi|\ll\sqrt{\Gamma_{2}^{2}+4\Delta^{2}}$ in the system under
consideration cooling mechanism is similar to the Doppler mechanism.
We will discuss it in more detail. The two-photon Raman transition
(the thick arrow on fig.\ref{fig:transit}) can occur in two ways:
with the virtual absorption of \(\pi\) and the emission of
\(\sigma^{+}\) lattice field components or with the absorption of
\(\sigma^{-}\) and the emission of \(\pi\) components. The
probability amplitudes of these processes are equal (see
(\ref{eq:complat})) and proportional \(\sin(\theta)\sin(2ky)\). The
probability amplitude under consideration contains the contributions
from two effective running waves ($\textbf{K}_{1}$ and
$\textbf{K}_{2}$ on fig. \ref{fig:dopl}) with the wave vectors
projections on the axis $y$ $K_{1}=2k$ and $K_{2}=-2k$, because
$\sin(2ky)$ is the superposition of two exponents $\exp(\pm2iky)$.
Let the effective detuning is negative. Then as atom moves towards
the wave $\textbf{K}_{1}$, its emission comes near the two-photon
resonance due to the Doppler effect, but the emission of the wave
$\textbf{K}_{2}$ comes far the resonance. So, the moving atom more
probably interacts with the contrepropagating effective wave, at
that it gets the momentum \(2\hbar k\). This process is two-photon,
that is the specificity of the cooling mechanism under consideration
relative to the standard Doppler mechanism. Other distinction is the
two-step relaxation of the exited state \(\mu_{g}=-1/2\), which is
characterized by the effective relaxation rate $\Gamma_{1}$.

When the Raman transitions is strong
$|\chi|\gg\sqrt{\Gamma_{2}^{2}+4\Delta^{2}}$ cooling mechanism is
similar to the Sisyphus mechanism \cite{cazan}, but it also has two
features. Firstly, the adiabatic potentials has the two times
shorter spatial period, secondly, the two-step transitions between
the adiabatic states are present, which are caused by the effective
relaxation.

\section{Kinetic coefficients in slow atom approximation}

In the general case of the atomic motion in nonuniform field the
kinetic coefficients \(F\) and \(D\) can be calculated by numerical
methods (for example, by the continuous fraction method
\cite{minogin}). In order to obtain the analytical expressions for
\(F\) and \(D\) one should use some approximations. The slow atom
approximation has a great importance (in particular, for the
temperature estimation). In our case it can be written as:
\begin{equation}
 k v << \gamma S_{p},\label{eq:slowatom}
\end{equation}
where $v$ is the atomic velocity. This condition means that atom
shifts over a distance far less than the light wave length during
the optical pumping time. In this limit (\ref{eq:slowatom}), to
describe the dissipative processes it is sufficient to consider only
the two first terms in the expansion of the force in velocity:
\begin{equation}
F(y,p_{y})\simeq F_{0}(y)+\alpha(y)v+....\label{eq:F}
\end{equation}
Here \(\alpha\) is the friction coefficient, \(F_{0}\) is the force
in the in zeroth order in velocity.For the diffusion coefficient we
take in to only the zeroth-order terms:
\begin{equation}
D(y,p_{y})\simeq D(y)=D_{sp}(y)+D_{ind}(y).\label{eq:D1}
\end{equation}
 The analytical expressions for the
coefficients \(F_{0}(y),\alpha(y),D(y)\) can be obtained by the
method of work \cite{forse}.

Let as demonstrate the results of analytical calculations for the
local magnitudes of the Fokker-Plank equation kinetic coefficients
in the slow atom approximation. It is convenient for brevity to use
the effective two-level system parameters (\ref{eq:effpar}) and
introduce the effective saturation parameter
\begin{equation}
S=\frac{\chi^{2}
\sin^{2}(2ky)}{\Gamma_{2}^{2}/2+\Delta^{2}}.\label{eq:suteff}
\end{equation}
The force in the zeroth order in velocity is
\begin{eqnarray}
F_{0}=4\hbar k \coth(\theta)\chi \sin(2ky)-\frac{4\hbar
k\Gamma_{1}\Delta S}{9(\Gamma_{1}+2\Gamma_{2}S)}.\label{eq:F0}
\end{eqnarray}
The friction coefficient is written as:
\begin{eqnarray}
\alpha=32\hbar k^{2} \Gamma_{1} \Delta
\left(\frac{\Gamma_{1}^{2}\Gamma_{2}S-
\Bigl[\Gamma_{2}^{3}+\Gamma_{1}\Gamma_{2}^{2}+4\Delta^{2}(\Gamma_{2}-\Gamma_{1})\Bigr]S^{2}}{\Gamma_{2}^{2}+4\Delta^{2}}-
2\Gamma_{2}S^{3}\right)/\Bigl(\Gamma_{1}+2\Gamma_{2}S\Bigr)^{3}\label{eq:alpha}
\end{eqnarray}
the induced diffusion coefficient has following form:
\begin{eqnarray}
D_{ind}=2\hbar^{2}k^{2}S\left(\Gamma_{1}^{3}\Gamma_{2}-
2\frac{\Gamma_{1}^{2}\Gamma_{2}\Bigl[-3\Gamma_{2}^{3}+4\Delta^{2}(4\Gamma_{1}-3\Gamma_{2})\Bigr]}{\Gamma_{2}^{2}+4\Delta^{2}}S-\right.\nonumber\\
\Biggl.-4\Gamma_{1}\Bigl[-3\Gamma_{2}^{3}+8\Delta^{2}(\Gamma_{1}-\Gamma_{2})\Bigr]S^{2}
+8(\Gamma_{2}^{2}+4\Delta^{2})\Gamma_{2}^{2}S^{3}\Biggr)/\Bigl(\Gamma_{1}+2\Gamma_{2}S\Bigr)^{3}\label{eq:difind}
\end{eqnarray}
the spontaneous diffusion coefficient is
\begin{equation}
D_{sp}=\frac{\hbar^{2}k^{2}\Gamma_{1}\Gamma_{2}S}{2(\Gamma_{1}+2\Gamma_{2}S)}\label{eq:difsp}
\end{equation}

We represent the saturation effective parameter $S$ as
$S=S_{0}\sin^{2}(2ky)$. In this case after the averaging on lattice
period the analytical expressions for kinetic coefficients have the
following form. The friction coefficient is written as:
\begin{eqnarray}
&<\alpha>=4 \hbar
k^{2}\Delta\Biggl(2\Gamma_{1}\Gamma_{2}^{2}S_{0}^{2}(\Gamma_{2}^{2}-4\Delta^{2})-2\Gamma_{1}^{3}(\Gamma_{2}^{2}+4\Delta^{2})-\Biggr.\nonumber\\
&-\Gamma_{2}^{3}S_{0}^{2}(\Gamma_{2}^{2}+4\Delta^{2})+2\Gamma_{1}^{5/2}\sqrt{\Gamma_{1}^{2}+2\Gamma_{2}S_{0}}(\Gamma_{2}^{2}+4\Delta^{2})+\nonumber\\
&\Biggl.+4\Gamma_{1}^{3/2}\Gamma_{2}S_{0}\sqrt{\Gamma_{1}+2\Gamma_{2}S_{0}}(\Gamma_{2}^{2}+4\Delta^{2})-2\Gamma_{1}^{2}\Gamma_{2}S_{0}(\Gamma_{2}^{2}+12\Delta^{2})\Biggr)/\nonumber\\
&\Biggl(\sqrt{\Gamma_{1}}\Gamma_{2}^{2}(\Gamma_{1}+2\Gamma_{2}S_{0})^{3/2}(\Gamma_{2}^{2}+4\Delta^{2})\Biggr).\label{eq:alphaaver}
\end{eqnarray}
The induced diffusion coefficient is
\begin{eqnarray}
&<D_{ind}>=\hbar^{2}k^{2}\Biggl(-\sqrt{\Gamma_{1}}\Gamma_{2}(6\Gamma_{1}^{2}+20\Gamma_{1}\Gamma_{2}S_{0}+\Biggr.\nonumber\\
&+15\Gamma_{2}^{2}S_{0}^{2})(\Gamma_{2}^{2}+4\Delta^{2})+2(\Gamma_{1}+2\Gamma_{2}S_{0})^{3/2}\nonumber\\
&(\Gamma_{2}^{4}S_{0}+8\Gamma_{1}^{2}\Delta^{2}+4\Gamma_{1}\Gamma_{2}\Delta^{2}+4\Gamma_{2}^{2}S_{0}\Delta^{2})+\nonumber\\
&+\sqrt{\Gamma_{1}}\Bigl\{-(2\Gamma_{1}^{2}+6\Gamma_{1}\Gamma_{2}S_{0}+3\Gamma_{2}^{2}S_{0}^{2})\Bigr.\nonumber\\
&(-3\Gamma_{2}^{3}+8(\Gamma_{1}-\Gamma_{2})\Delta^{2})+\Bigl[2\Gamma_{2}^{4}S_{0}\Bigl(3\Gamma_{2}S_{0}(\Gamma_{2}^{2}+4\Delta^{2})+\Bigr.\nonumber\\
&\Bigl.\Bigl.\Bigl.\Gamma_{1}(\Gamma_{2}^{2}+4(1-2S_{0})\Delta^{2})\Bigr)\Bigr]/(\Gamma_{2}^{2}+4\Delta^{2})\Bigr\}\Biggr)/\Biggl(2\Gamma_{2}^{3}(\Gamma_{1}+2\Gamma_{2}S_{0})^{3/2}\Biggr)
\end{eqnarray}
The spontaneous diffusion coefficient has following form:
\begin{eqnarray}
<D_{sp}>=\frac{\hbar^{2}k^{2}\Gamma_{1}}{4}\Biggl(-1+\sqrt{\frac{\Gamma_{1}+2\Gamma_{2}S_{0}}{\Gamma_{1}}}\Biggr).\label{eq:Dspysr}
\end{eqnarray}
Note that at $\Gamma_{1}=\Gamma_{2}$, our expressions formally
coincide (with an accuracy of constant factors) with the
corresponding formulas for two-level atom in a standing wave field
\cite{ashkin,cazan}.
\section{Discussion of the results}
We estimate the kinetic temperature by the standard way
\cite{cazan},
 neglecting the spatial localization:
\begin{equation}
k_{B}T=-\frac{\left<D_{ind}\right>+\left<D_{sp}\right>}{\left<\alpha\right>},\label{eq:temp}
\end{equation}
where $k_{B}$ is the Boltcman constant. In the weak Raman
transitions limit expression (\ref{eq:temp}) is written as:
\begin{equation}
k_{B}T=-\frac{5\hbar\left((\frac{\Gamma_{2}}{2})^{2}+\Delta^{2}\right)}{16\Delta}.\label{eq:tempdop}
\end{equation}
In this limit the minimal temperature is achieved at the effective
detuning \(\Delta=-\frac{\Gamma_{2}}{2}\), and it is equal
\(k_{B}T=\frac{5}{16}\hbar\Gamma_{2}\). In our case it is possible
to change the minimal temperatures by variation the effective
relaxation constant \(\Gamma_{2}\) (for this purpose, it is
necessary to change the parameters of \(\Gamma_{2}\), for example,
the pumping field intensity). This feature presents the important
difference from the usual Doppler cooling in two-level system.

\begin{figure}[t]
\centering
\includegraphics[height=6cm]{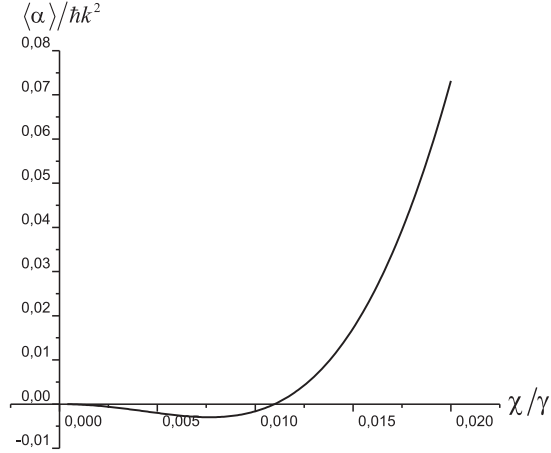}
\caption{Dependence of averaged friction coefficient on the Rabi
frequency (the effective detuning \(\Delta=-0.1\gamma\)
(\(\delta_{p}=0,\Omega_{p}=0.1\gamma,\omega_{z}=0.1\gamma\)).}\label{fig:alpha}
\end{figure}

Further, let us consider the dependence of averaged friction
coefficient and the atomic temperature on the effective Rabi
frequency. In fig. \ref{fig:alpha} the dependence of averaged
friction coefficient on the effective Rabi frequency is presented
(the effective detuning \(\Delta=-0.1\gamma\)).
It is clearly from fig. \ref{fig:alpha} that in the weak Raman
transition limit(when the effective Rabi frequrncy \(\chi\) is
small) \(\left<\alpha\right><0\), that is the cooling of atoms
occurs. In the strong Raman transition limit(when \(\chi\) is large)
\(\left<\alpha\right>>0\), that is the heating of atoms occurs. This
dependence of kinetic processes  direction on the effective Rabi
frequency qualitatively coincide with the form of analogically
dependence in two-level system.

\begin{figure}[h]
\centering
\includegraphics[height=4cm]{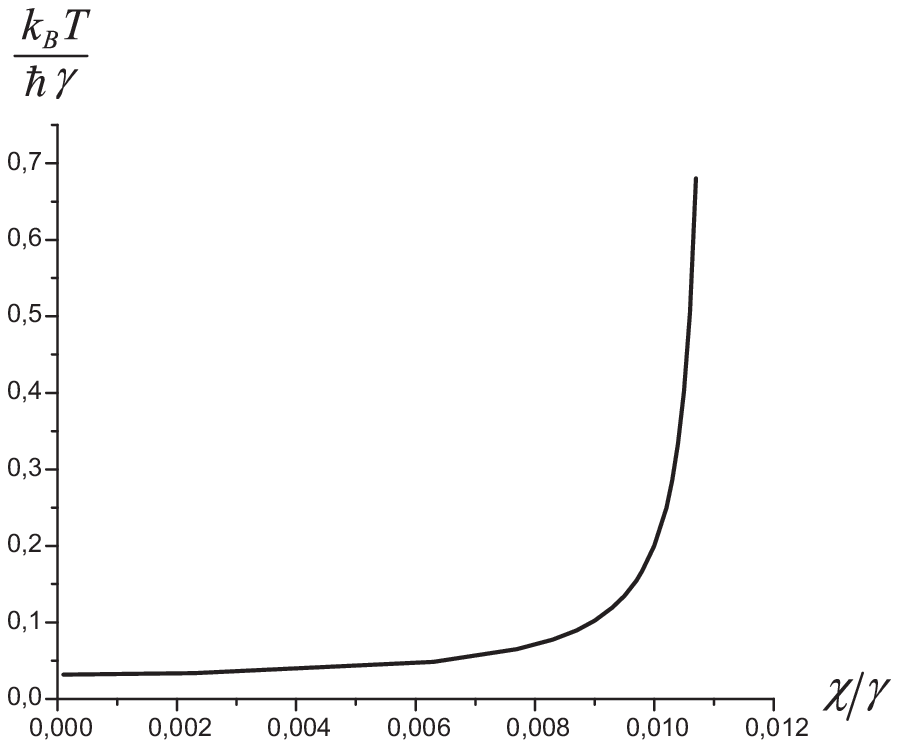}
\caption{Temperature dependence on Rabi frequency when
\(\Delta=-0.1\gamma\).}\label{fig:tempdop}
\end{figure}

\begin{figure}[h]
\centering
\includegraphics[height=4cm]{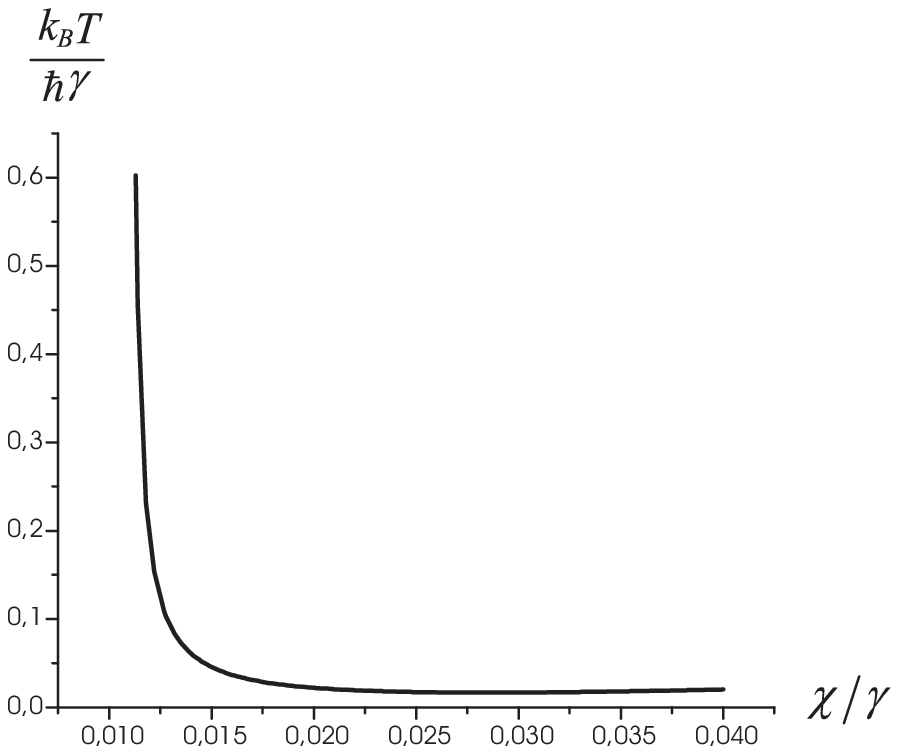}
\caption{Temperature dependence on Rabi frequency when
\(\Delta=0.1\gamma\)}\label{fig:tempsiz}
\end{figure}

On fig.\ref{fig:tempdop} the dependence of kinetic temperature
(\ref{eq:temp}) on \(\chi\) is presented (when
\(\Delta=-0.1\gamma\)).
This figure illustrates the cooling mechanism in the weak Raman
transitions limit. The atomic temperature decrease is observed as
\(\chi\) decreases, that corresponds to the Doppler cooling limit in
two-level system. When the effective detuning is positive cooling is
observed in the strong Raman transitions limit (fig.
\ref{fig:tempsiz}).
In the two-level system it corresponds to Sisyphus  cooling
mechanism. It is obviously from fig.\ref{fig:tempsiz} that the
temperature increase without limit as $\chi$ increase.

So, the demonstrated dependencies \(\left<\alpha\right>\) and
$\left<kT\right>$ on \(\chi\) confirm the qualitative interpretation
of cooling mechanisms given above.
\section{Comparison with experiment}
The previously derived theoretical results can be compared to the
experimental data \cite{swid}. For this purpose let as calculate the
model parameters that are correspond to the experimental conditions:
the pumping field detuning +2.3 MHz,the pumping field intensity
\(I_{p}=0.24\, \mbox{mW}\,\mbox{cm}^{-2}\);the lattice field
detuning -9 GHz, the single-beam lattice intensity \(I_{L}=75\,
\mbox{mW} \,\mbox{cm}^{-2}\),the angle between lattice beams
polarization vectors \(45^{\circ}\);the magnetic field changed in
the range from 0 to 200 mG. The pumping and lattice field Rabi
frequencies are calculated from the formula \(\Omega_{p,L}=\gamma
\sqrt{I_{p,L}/(8 I_{s})}\), where \(I_{s}=1.1 \,\mbox{mWt}
\,\mbox{cm}^{-2}\) is saturation intensity for the $D_2$ line
$^{133}Cs$, $\gamma = 2 \pi \times$ 5.3 MHz. At the calculation of
Zeeman shift we use the $g$-factor value for lowest hyperfine level
of $^{133}Cs$ ground state: $g=-1/4$ that gives $\omega_z = 2\pi
\times 350 \,\mbox{kHz}\, \mbox{G}^{-1} B$. Under this conditions
the effective Rabi frequency is $\chi = 2 \pi \times 11$ kHz, and
the effective relaxation rates of two-level system are $ \Gamma_1 =
2\pi\times 76$ kHz and $\Gamma_2 = 2\pi\times 222$ kHz.
It is necessary to compare the magnitudes $|\chi|$ and
$\sqrt{\Gamma_{2}^{2}+4\Delta^{2}}$ for definition dominating
cooling mechanism under the given conditions. As the the magnetic
field changes in the range from 0 to 200 mG (that corresponds to the
experimental conditions) $\sqrt{\Gamma_{2}^{2}+4\Delta^{2}}$ changes
in the range from 0.29 MHz to 0.4 MHz. As this take place, $|\chi|$
remains 26 - 36 times smaller than this magnitude. Consequently, the
domain with mainly weak Raman connection and the Doppler-like
cooling mechanism correspond to the given conditions. According to
(\ref{eq:tempdop}) at the weak Raman coupling the minimal
temperature is achieved under the condition $\Delta = - \Gamma_2/2$
(that corresponds to $B_{min}=50$ mG). This minimal temperature can
be estimated as $T_{min}= 0.3 \,\hbar \Gamma_2/k_B \simeq 3.3$
$\mu$K. These values are close to the experimentally observed
\cite{swid} ($B_{min}=45$ G è $T_{min}=1.5$ $\mu$ K). The
experimentally obtained temperature is more than 2 times smaller
than the theoretical limit. This discrepancy is most likely due to
the disregarding of the contribution of atoms, confined to the
optical potential minima. It is necessary for the more detailed
analysis to consider simultaneously the cooling of unbound and bound
atoms. This consideration is beyond the purpose of the present work
and will appear the subject of further investigations.

Moreover, we express the dependence of the kinetic temperature on
the magnetic field magnitude (fig.7) (It was calculated from the
formula (\ref{eq:temp}) that is with the taking into account all
orders on $|\chi|$).
\begin{figure}[h]
\centering
\includegraphics[height=4cm]{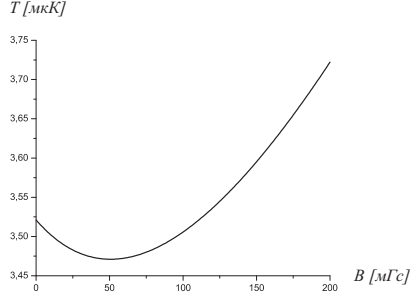}
\caption{Temperature dependence on magnetic field under
\(\delta_{p}=2 \pi \times 2.3 \, \mbox{MHz}, \,\delta_{L}=-2 \pi
\times 9 \, \mbox{MHz}, I_{p}=0.24\, \mbox{mWt}
\,\mbox{cm}^{-2},I_{L}=75\, \mbox{mWt} \,\mbox{cm}^{-2},
\theta=45^{\circ} \)}
\end{figure}
It is clear from the comparison fig.7 and the experimental
dependence of atomic temperature on magnetic field (work
\cite{swid}, fig. 6) that there is a satisfactory qualitative
agreement between the results of our theoretical model and the
experimental data.
\section{Conclusion}
Let as summarize some results. We considered the laser cooling of
the unbound atoms with the exited state and ground state momentum
\(J_{e}\) and \(J_{g}\) that were equal \(1/2\) in one-dimensional
\textit{lin-\(\theta\)-lin} lattice field configuration. It was
showed that in the low saturation limit in pumping field
(\ref{eq:sut}) the qualitative interpretation of cooling mechanisms
could be made in the framework of the consideration of the effective
two-level system which was formed by the ground-state sublevels. We
compared the equations, that described the effective system with the
known equations for two-level system. The dependence of effective
parameters on model parameters was found from this comparison. The
qualitative interpretation of cooling mechanisms was given. It was
showed that in the weak Raman transitions limit the Doppler-like
mechanism was observed, and in the strong Raman transitions limit
the similar Sisyphus mechanism was demonstrated. The analytical
expressions for the force acting on atom, spontaneous and induced
diffusion coefficients were obtained. The quantitative estimate of
atomic kinetic temperature was made. It was demonstrated that the
dependence of friction coefficient and temperature on the effective
Rabi frequency confirmed our qualitative interpretation of cooling
mechanisms. The comparison of theoretical calculations of
temperature and experimental data of work \cite{swid} was made and a
satisfactory qualitative agreement was revealed. The results of this
work can be used for analysis of laser cooling of atoms in
nondissipative optical lattices.

This work was supported by RFBR (05-02-17086, 05-08-01389,
07-02-01230, 07-02-01028), INTAS-SBRAS (06-1000013-9427) and by
Presidium of Siberian Branch of Russian Academy of Sciences. N.A.M.
was also supported by "Dynasty" Fund.

\end{document}